\documentclass[preprint,floatfix] {revtex4} 
 
\usepackage{graphicx}
\usepackage{subfigure}
\begin{document}

\title{Studies on some exponential-screened Coulomb potentials}
\author{Amlan K. Roy}
\altaffiliation{Email: akroy@iiserkol.ac.in, akroy6k@gmail.com}
\affiliation{Division of Chemical Sciences,   
Indian Institute of Science Education and Research (IISER)-Kolkata, 
Mohanpur Campus, P. O. BCKV Campus Main Office, Nadia, 741252, WB, India}

\begin{abstract}
The generalized pseudospectral method is employed to study the bound-state spectra of some of the 
exponentially screened Coulomb potentials, \emph{viz.}, the exponential cosine screened Coulomb (ECSC) 
and general exponential screened Coulomb (GESC) potential, with special emphasis on \emph{higher} states and 
\emph{stronger} interaction. Eigenvalues accurate up to eleven significant figures are obtained through a 
non-uniform optimal spatial discretization of the radial Schr\"odinger equation. All the 55 eigenstates 
of ECSC potential with $n \le 10$ and 36 eigenstates of GESC potential with $n \le 8$ are considered for 
arbitrary values of the screening parameter, covering a wide range of interaction. Excited states as high
as up to $n=18$ have been computed with high accuracy for the first time. Excellent agreement with the 
literature data has been observed in all cases. All the GESC eigenstates are calculated with much greater 
accuracy than the existing methods available in literature. Many $l \ne 0$ states of this potential 
are reported here. 
In both cases, a detailed variation of energies with respect to the parameters in potential is monitored. 

\end{abstract}
\maketitle

\section{Introduction}
Realistic potentials which describe quantum mechanical systems, are not usually exactly solvable 
in the Schr\"odinger picture, except for a few occasions such as Harmonic oscillator, Coulomb 
potential, etc. Therefore, finding exact analytical solution of Schr\"odinger equation for a 
given potential corresponding to a physical system of interest, constitutes one of the major 
challenges in quantum mechanics. This is a common problem, and often encountered in almost every 
branches, such as atomic, molecular, solid-state, nuclear, particle and plasma physics, etc. A 
large number of attractive promising approximate formalisms have been developed ever since the 
inception of theory, which can provide highly accurate or \emph{near-exact} results in some 
cases. However, the same for a \emph{general} potential for \emph{any} allowed values of quantum 
numbers, for \emph{arbitrary} values of potential parameters (if present in the system) still 
remains elusive, and thus always has been an active area of research.

Here we are concerned with the accurate bound states of two central singular potentials, namely 
(i) the generalized exponential cosine screened Coulomb potential, given by, 
\begin{equation}
v(r)=-\frac{A}{r} \ e^{-\delta_1r} \ \cos(g \delta_2 r)
=-\frac{A}{r} \ e^{-\delta r} \ \cos (g \delta r) \ \ \ 
(\mathrm{when \ \delta_1 = \delta_2 = \delta}),
\end{equation}
and (ii) a much less frequently studied, general exponential screened Coulomb (GESC) potential of 
the form, 
\begin{equation}
v(r)=-\frac{a}{r} \ [1+(1+br) e^{-2br}].
\end{equation}
In Eq.~(1), $A$ represents the coupling strength constant while $\delta_1, \delta_2$ are two 
screening parameters. This potential reduces to the familiar Yukawa potential for $g=0$, which has 
wide applications in nuclear, solid-state and plasma physics. For $g=1$, this is termed as the
exponential cosine screened Coulomb (ECSC) potential, and it is in this form that this potential has 
been studied maximum. One distinctive feature of this oscillating potential (in contrast to the 
Coulomb potential) is the \emph{finite} number of bound states, i.e., such 
states exist only for certain values of the screening parameter below a threshold limit (the 
so-called critical $\delta_c$). In other words, the total number of different 
energy levels is finite for a given value of $\delta > 0$. Similarly, the two potential parameters 
$a,b$ in Eq.~(2) signify coupling strength and screening parameters respectively. Throughout the whole 
article, $a$ is fixed at unity. These two potentials 
can be used to represent the effective interaction in many-electron atoms; also they have important 
applications in solid-state, nuclear and plasma physics as well as in field theory \cite{anderson52,kubo52,
bonch59,propokev67,ferrell74,brezin79,weisbuch93,harrison00,shukla08}. Lately, the effect of screening
on atomic photoionization in H and He$^+$ has been studied by means of Yukawa and ECSC potential 
\cite{lin10}. Also, the ground and excited resonances in two-electron systems such as He, molecular 
H$_2$ in ECSC as well as generalized screened potential have been investigated 
\cite{ghoshal09,ghoshal09a,ghoshal11,ghoshal11a}.

None of these potentials admits exact analytical result. Therefore, over five decades, a 
considerably large number of attempts have been made to calculate their eigen spectra accurately. 
Here we mention a few of them. Perturbation and variational methods were used \cite{lam72} to 
produce eigenvalues with reasonable accuracy, as well as the number of bound states for a given 
value of $\delta$. The $s$ states were reported \cite{dutt79,ray80} via the representation of ECSC 
potential by a Hulth\'en potential with an energy-dependent strength parameter, through the use of 
an Ecker-Weizel approximation. The $n \le 4$ states have been calculated by means of a hyper-virial
Pad\'e approximation \cite{lai82}, dynamical group approach \cite{meyer85}, hyper-virial equation 
with Hellman-Feynman theorem \cite{sever90}, etc. Using a numerical method  \cite{singh83}, all the 
36 states below $n \le 8$, as well as the critical screening parameters were obtained within an 
accuracy of eight to six significant figures. Analytical expressions for eigenvalues and 
eigenfunctions of ground and first excited states, up to 14 terms, were presented by a large-$N$ 
expansion method \cite{sever87}. Eigenvalues of $1s$ to $8k$ states have been reported by means of 
a shifted $1/N$ expansion technique \cite{ikhdair93}. An iterative solution for eigenvalues 
belonging to arbitrary $n,l$ quantum numbers has been put forth by employing an asymptotic 
iteration method \cite{bayrak07}. A novel perturbation method \cite{ikhdair07} has also been 
proposed for this potential where the radial Schr\"odinger equation is decomposed into two parts, 
one of them being exactly solvable while the other part leading to closed analytical solution or 
an approximate treatment depending on the potential in question. Lately, a J-matrix approach 
\cite{nasser11} with a Gaussian quadrature scheme has offered high-quality results for bound and 
continuum states. A Ritz variation method with hydrogenic wave function as the trial function 
\cite{paul11} has also produced promising results for such potential. Recently, an analytical 
scheme \cite{bahlouli12} inspired by the J-matrix method, has been quite successful for such 
potentials. The $\delta_c$ values have been 
estimated by numerical \cite{lam72,singh83,ikhdair93,nasser11}, as well as analytical methods 
\cite{dutt80}. 

The GESC potential, on the other hand, has not received much attention. I am aware of only two 
studies. Energy eigenvalues of ground and first excited states were presented up to 14 terms using 
a large-$N$ expansion method \cite{sever87a}. In another attempt \cite{ikhdair07a}, the 
perturbative method of \cite{ikhdair07} was used to obtain the $1s, 2s, 3s$ states of this 
potential with decent accuracy. 

In this work, we study the eigenspectra of both these potentials in Eqs.~(1) and (2) using a 
generalized pseudospectral (GPS) method, which has been quite successful for a variety of systems 
such as the spiked harmonic oscillator, Hulth\'en and Yukawa potentials, power-law and logarithmic 
potentials, ground and excited states (low- and high-lying Rydberg states) of atoms as well as 
other singular systems \cite{roy04, 
roy04b,roy05,roy05a,roy07,roy08a,roy11}. Potential parameters are scanned over a large domain. In 
few occasions, for some of the methodologies mentioned above for ECSC potential, it so happens that, 
eigenvalues and eigenfunctions are quite difficult to calculate for high-lying states and also at 
certain region of the potential parameter, especially near the $\delta_c$. Here we pay special 
attention to both these issues for a better understanding of their spectra and also to judge the 
validity and efficacy of the method. To this end, 
accurate energies and wave functions are presented for all the 55 levels belonging to $n=10$ 
states of ECSC potential to extend the domain of applicability of the GPS procedure. Variation of 
the same with respect to $\delta$ are also monitored. For the GESC potential, only some low-lying
$l=0$ states have been considered so far in the literature; no results are available for $l \ne 0$ 
states. So here, we report all the states up to $n=8$ for a wide range of parameters in the 
potential for the first time, and some higher states as well. A detailed comparison with the 
available results in the 
literature has been made, wherever possible. The article is organized as follows. An outline of the
theory and method of calculation is presented in Section II. A discussion of the results is given 
in Section III, while a few concluding remarks are made in Section IV.  

\section{The GPS method}
\label{sec:method}
The section gives the essential steps of GPS approach, as implemented here for the solution of 
single-particle Schr\"odinger equation for a non-relativistic Hamiltonian containing an 
exponentially screened potential term. The key advantage of the approach is that it offers a 
\emph{non-uniform, optimal} spatial discretization. That means one can use a finer grid at small $r$ 
and coarser grid at large $r$, maintaining high-accuracy at both these regions. This also implies only
a small number of spatial points suffices to achieve convergence. Thus compared to standard finite 
difference/finite element methods, the GPS scheme is both accurate and efficient. Other details 
could be found in the references \cite{roy04, roy04b, roy05, roy05a, roy07, roy08, roy08a, roy11}. 
Unless otherwise mentioned, atomic unit is used throughout the article. 

The radial Schr\"odinger equation can be written in the following working form, 
\begin{equation}
\left[-\frac{1}{2} \ \frac{\mathrm{d^2}}{\mathrm{d}r^2} + \frac{\ell (\ell+1)} {2r^2}
+v(r) \right] R_{n,\ell}(r) = E_{n,\ell}\ R_{n,\ell}(r)
\end{equation}
where $v(r)$ is as given in Eq.~(1) or (2), whereas $n$, $\ell$ signify the usual radial and angular 
momentum quantum numbers respectively. 

As a key step, a function $f(x)$, defined in the interval $x \in [-1,1]$, is approximated by the N-th 
order polynomial, $f_N(x)$, through a cardinal function $g_j(x)$, as follows: 
\begin{equation}
f(x) \cong f_N(x) = \sum_{j=0}^{N} f(x_j)\ g_j(x).
\end{equation}
This guarantees that the approximation is \emph {exact} at the \emph {collocation points} $x_j$, 
i.e., $ f_N(x_j) = f(x_j)$, and requires that the cardinal function satisfies 
$g_j(x_{j'})=\delta_{j'j}$. Here we use the Legendre pseudospectral method, where $x_0=-1$, 
$x_N=1$, and the $x_j (j=1,\ldots,N-1)$ are obtained from roots of first derivatives of the Legendre 
polynomial, $P_N(x)$ with respect to $x$, as $ P'_N(x_j) = 0$.
The $g_j(x)$ are given by,
\begin{equation}
g_j(x) = -\frac{1}{N(N+1)P_N(x_j)}\ \  \frac{(1-x^2)\ P'_N(x)}{x-x_j},
\end{equation}
Now, the semi-infinite domain $r \in [0, \infty]$ is mapped onto a finite domain $x \in [-1,1]$ 
via the transformation $r=r(x)$. At this stage, one could introduce an algebraic nonlinear mapping
of the form, 
\begin{equation} 
r=r(x)=L\ \frac{1+x}{1-x+\alpha},
\end{equation}
with L and $\alpha=2L/r_{max}$ as two mapping parameters, to obtain a transformed differential 
equation as: $ f(x)=R_{nl}(r(x))/\sqrt{r'(x)}$. Now, one applies the Legendre pseudospectral method 
to this equation and finally a symmetrization procedure to yield the following \emph{symmetric} 
eigenvalue problem,
\begin{equation}
\sum_{j=1}^{N-1} \left[ -\frac{1}{2}D_{ij}+u_j\delta_{ij} \right] \chi_j = \epsilon_{nl} \chi_i .
\end{equation}
This is readily solved by standard available routines such as that in NAG Fortran library, giving 
highly accurate eigenvalues and eigenfunctions. After some straightforward algebra, one finds that,  
\begin{equation}
\chi_i\!= \! R_{nl}(r_i) \ \sqrt{(r'_i)}/P_N(x_i), \ \ \ u_i\!=\!l(l+1)/2r_i^2 
+v(r_i), 
\end{equation}
with $\chi_i\!=\!\chi(x_i), u_i\!=\!u(x_i), r_i\!=\!r(x_i), r'_i\!=r'(x_i)$, while $D_{ij}$ 
denotes the symmetrized second derivative of cardinal function, given as follows,  
\begin{eqnarray}
D_{ij} & = & -\frac{2}{r_i'(x_i-x_j)^2r'_j}, \  \ \ \ i \neq j, \nonumber \\
& = & -\frac{N(N+1)}{3r_i'^2(1-x_i^2)}, \ \ \ i= j.
\end{eqnarray}
A series of calculation was performed for various potential parameters with respect to the grid 
mapping parameters to ascertain the accuracy and reliability of the current method. In this way, 
a ``stable" grid was found, which appears to be sufficient for all the converged results presented 
in this article. Unless otherwise mentioned, all the reported results correspond to this consistent 
set of parameters, $\alpha=25, N=200$ and $r_{max}=300$. There are some instances, where this set is
not adequate, and appropriate variations are allowed; these are mentioned appropriately in the text. 
Current results are reported only up to the precision that maintained stability. Eigenvalues are 
{\em truncated} rather than {\em rounded-off}, and hence may be considered as correct up to all the 
decimal places they are reported.

At this stage, a few remarks may be made regarding the GPS method. Typically in direct numerical 
methods, one truncates the semi-infinite domain into a finite domain $[\mathrm{r_{min}, r_{max}}]$ 
to deal with the problems of singularity at $r=0$, and infinite domain. In order for this, 
$\mathrm{r_{min}}$ and $\mathrm{r_{max}}$
need to be chosen sufficiently small and large respectively. This consequently results in a 
rather large number
of grid points and also, in general, introduces some truncation error. To overcome this problem, 
one can map the semi-infinite domain $[0,\infty]$ exactly into the finite domain [$-$1,1] using the 
mapping $r=f(x)$ (Eq.~6) so that the Legendre pseudopotential technique can be applied. This 
introduces an additional undesirable feature; namely it leads to an unsymmetric or generalized 
eigenvalue problem, which in turn, is bypassed via the symmetrization procedure mentioned above. 
The method has been successfully applied to resonance states as well. For these and many other 
features of the method, the interested reader is referred to the references 
\cite{yao93,wang94,telnov99} and those therein.

\section{Results and Discussion}

\begingroup
\squeezetable
\begin{table}
\caption {\label{tab:table1}Calculated negative eigenvalues E (in a.u.) of some
selected $s$ states of the ECSC potential for various $\delta$ along with the
literature data. The $\delta_c$ values of $1s, 2s, 3s, 4s, 8s$ are 0.72, 
0.16656, 0.0724, 0.0404 and 0.01 respectively, taken from \cite{nasser11}.} 
\begin{ruledtabular}
\begin{tabular}{cclllll}
State & $\delta$ & \multicolumn{2}{c}{$-$Energy}    & $\delta$ 
& \multicolumn{2}{c}{$-$Energy} \\ 
\cline{3-4} \cline{6-7}
    &  & This work   & Reference &   &  This work & Reference \\   \hline
$1s$  & 0.0002 & 0.49980000000 & 0.499800\footnotemark[1]$^,$\footnotemark[2]$^,$\footnotemark[3], 
                                 0.499899\footnotemark[4]  
      & 0.06   & 0.44020051029 & 0.440200\footnotemark[1]$^,$\footnotemark[8]$^,$\footnotemark[11],
                                 0.440201\footnotemark[2]$^,$\footnotemark[3]$^,$\footnotemark[5], \\
      &        &   &   &   &  &  0.44020051\footnotemark[6]$^,$\footnotemark[10],
                                 0.4402004\footnotemark[7],                           \\
      &        &   &   &   &  &  0.44020057\footnotemark[9],0.44020051029\footnotemark[12]  \\
$1s$  & 0.1    & 0.40088477464 & 0.400883\footnotemark[1],0.400884\footnotemark[2],
                                 0.400885\footnotemark[3]$^,$\footnotemark[5]$^,$\footnotemark[13],   
      & 0.7    & 0.00115044274 & $-$0.050624\footnotemark[1],$-$0.036908\footnotemark[2],  \\
      &        &              &  0.402155\footnotemark[4],0.40088477\footnotemark[6],
                                 0.4008839\footnotemark[7]$^,$\footnotemark[8],    
      &        &              &  0.00184\footnotemark[3],$-$0.000043\footnotemark[5],   \\
      &        &              &  0.40088421\footnotemark[9],0.40088476\footnotemark[10],   
      &        &              &  0.00115044272\footnotemark[12]  \\
      &        &              &  0.4008785\footnotemark[11],0.400884774639\footnotemark[12]    
      &        &              &   \\
$2s$  & 0.06   & 0.06742110520 & 0.067385\footnotemark[1],0.067408\footnotemark[2],
                                 0.067421\footnotemark[3]$^,$\footnotemark[5], 
      & 0.165  & 0.00018502068 & 0.0001850\footnotemark[12]     \\
      &        &              &  0.067525\footnotemark[4],
                                 0.0674217\footnotemark[7]$^,$\footnotemark[8],
                                 0.06742608\footnotemark[9],       &        &              &   \\
      &        &              & 0.06742085\footnotemark[10],0.0673900\footnotemark[11],    
      &        &              &   \\
      &        &              & 0.06742110514\footnotemark[12],0.06742173\footnotemark[14]    
      &        &              &   \\
$3s$  & 0.04   & 0.01882306336 & 0.018707\footnotemark[1],0.018768\footnotemark[2],
                                 0.018822\footnotemark[3], 
      & 0.072  & 0.00009790825 & 0.0000979\footnotemark[12]  \\ 
      &        &              & 0.019604\footnotemark[4],0.018823\footnotemark[5],
                                0.0188478\footnotemark[7],       &        &              &   \\
      &        &              & 0.01886716\footnotemark[9],0.018821\footnotemark[10],     
      &        &              &   \\
      &        &              & 0.0188586\footnotemark[11],0.01882306333\footnotemark[12]     
      &        &              &   \\
$4s$  & 0.0005 & 0.03075002676 & 0.030750\footnotemark[1]$^,$\footnotemark[2]$^,$\footnotemark[3],
                                 0.030751\footnotemark[4]    
      & 0.005  & 0.02627512430 & 0.026275\footnotemark[1]$^,$\footnotemark[2]$^,$\footnotemark[3],
                                 0.026321\footnotemark[4]        \\
$4s$  & 0.02   & 0.01257177727 & 0.012539\footnotemark[1],0.012557\footnotemark[2],
                 0.012572\footnotemark[3]$^,$\footnotemark[5]$^,$\footnotemark[10],   
      & 0.04   & 0.00014026953 & $-$0.001079\footnotemark[1],$-$0.000670\footnotemark[2],  \\
      &        &              & 0.013084\footnotemark[4],0.0125811\footnotemark[7],
                                0.01259233\footnotemark[9]  
      &    &   & 0.000118\footnotemark[3],0.000125\footnotemark[5],                    \\
      &        &              &                                                                         
      &    &   & 0.0010694\footnotemark[7],0.00032273\footnotemark[9]              \\
$8s$  & 0.0001 & 0.00771250340 & 0.007713\footnotemark[13]    & 0.005  & 0.00314139349 & 
                                 0.003134\footnotemark[13]              \\
$17s$ & 0.0005 & 0.00123778635 &          & 0.001  & 0.00078491417 &              \\
$18s$ & 0.0005 & 0.00105272559 &          & 0.001  & 0.00061036597 &              \\
\end{tabular}
\end{ruledtabular}
\begin{tabbing}
$^{\mathrm{a}}$Perturbation (Coulomb), Ref.~\cite{lam72}. \hspace{30pt}  \= 
$^{\mathrm{b}}$Perturbation (Hulthen), Ref.~\cite{lam72}. \hspace{30pt}   \=
$^{\mathrm{c}}$One-parameter variational, Ref.~\cite{lam72}. \hspace{30pt}  \\
$^{\mathrm{d}}${Ref.~\cite{ray80}.}  \hspace{42pt} \=
$^{\mathrm{e}}${Ref.~\cite{lai82}.}  \hspace{42pt} \=
$^{\mathrm{f}}${Ref.~\cite{singh83}.}   \hspace{42pt} \=
$^{\mathrm{g}}${Ref.~\cite{meyer85}.}  \hspace{42pt} \= 
$^{\mathrm{h}}${Ref.~\cite{sever87}.}  \hspace{42pt} \= 
$^{\mathrm{i}}${Ref.~\cite{ikhdair93}.}  \hspace{42pt}  \=    \\ 
$^{\mathrm{j}}${Ref.~\cite{bayrak07}.}  \hspace{42pt}  \=          
$^{\mathrm{k}}${Ref.~\cite{ikhdair07}.}  \hspace{42pt}  \=           
$^{\mathrm{l}}${Ref.~\cite{nasser11}.}  \hspace{38pt}  \=
$^{\mathrm{m}}$Two-parameter variational, Ref.~\cite{lam72}. \hspace{20pt}  \=
$^{\mathrm{n}}${Ref.~\cite{paul11}.}  \\
\end{tabbing}
\end{table}

At first, in Table I, we report some $s$ states of the ECSC potential for low as well as high excitations. 
For all the ECSC potential calculations throughout the article, parameter $A$ is set to unity. 
A wide range of screening parameters is considered--including low, intermediate and high values, 
signifying small, intermediate and large interaction respectively. Critical values of the screening 
parameter, taken from \cite{nasser11}, are also mentioned in the table for $n \le 8$ for easy understanding. 
As evident, a large number of results are available in the literature for comparison, which we quote 
accordingly. One of the very first definitive calculations of this potential was reported in 
\cite{lam72}. All the $s$ states considered here with $n$=1,4 (except $2s, 3s$ in the high-screening region) 
were estimated by first-order perturbation treatment with (a) Coulomb potential as unperturbed potential 
(b) Hulth\'en potential as unperturbed potential, and (c) a one-parameter variational calculation with 
reasonably good accuracy. An Ecker-Weizel approach has been used for 1$s$--4$s$ states in the 
low-screening region through an approximation of the ECSC potential by Hulth\'en potential with modest 
accuracy \cite{ray80}. Both $s$ and $l \ne 0$ states of $n \leq 4$ were treated in intermediate 
coupling region within the hyper-virial Pad\'e approximation \cite{lai82}, a dynamical group approach 
\cite{meyer85}, and an asymptotic iteration method \cite{bayrak07}. The ground and first excited states 
of the ECSC potential for medium values of screening parameter have been obtained within the large-$N$ 
expansion method \cite{sever87} as well. Energies correct up to six to eight significant figures were 
reported by means of a shifted $1/N$ expansion \cite{ikhdair93}, for both $l=0$ as well as $l \ne 0$ 
states. Lately, a new perturbative scheme \cite{ikhdair07} has been put forth for the $n \le 3$ states 
with decent success. However, it seems that, so far the most accurate eigenvalues are reported by a 
J-matrix method \cite{nasser11}. In the neighborhood of low and moderate coupling, the present results 
are of very similar accuracy as those from \cite{nasser11} (in many occasions they coincide; otherwise 
they differ in the 11th or 12th place of decimal). In some states, for $\delta$s near the threshold 
limit, their results were reported for somewhat lesser accuracy. Through the 
present method, we are able to obtain eigenvalues of consistently better accuracy near the 
\emph{strong-coupling region} (see, for example, $2s$ and $3s$ for $\delta = 0.165$, 0.072 respectively). 
For higher-lying states, the reference literature values dramatically reduce in number so much so that 
for $n=8$, our results could only be compared with the lone two-parameter variational calculation 
\cite{lam72}, where the present results are visibly improved. And for states with $n >8$, no results 
could be found, and we report some sample results for $17s$ and $18s$ to emphasize the ease and ability 
of our method for higher states. Note that, for these higher states, however, the $r_{max}$ value needs 
to be suitably increased to achieve the desired convergence, e.g., for $17s, 18s$, an $r_{max}=1100$ a.u.,
was used, whereas the other two parameters $\alpha$ and $N$ needed no adjustments. 
This is reminiscent of a situation encountered earlier for Hulth\'en and Yukawa potentials \cite{roy05a}
in the stronger coupling regime. Some numerical results are also 
available for the low-lying states \cite{singh83,paul11}, which have been quoted as well. 

\endgroup
\begingroup
\squeezetable
\begin{table}
\caption {\label{tab:table2}Calculated negative eigenvalues (a.u.) of the ECSC potential for selected 
$n=2-6, \ell \neq 0$ states for various $\delta$ values along with the literature data. The $\delta_c$ 
values for $p,d,f,g,h$ states for $n=(l+1)$ up to 6 are as follows \cite{nasser11}: $np$: 0.1482, 
0.0687, 0.03926, 0.0253, 0.01765; $nd$: 0.06358, 0.037405, 0.0245, 0.01724; $nf$: 0.03524, 0.02348, 
0.016708; $ng$: 0.02237, 0.016099; $nh$: 0.015455.} 
\begin{ruledtabular}
\begin{tabular}{clllclll}
State & $\delta$ & \multicolumn{2}{c}{$-$Energy} & State &  $\delta$ &  
\multicolumn{2}{c}{$-$Energy} \\ 
\cline{3-4} \cline{7-8}
   &  & This work   & Literature &  &   &  This work & Literature \\   \hline
$2p$ &  0.002  & 0.12300007948 & 0.123000\footnotemark[1]   &
$5d$ &  0.005  & 0.01505072772 & 0.015051\footnotemark[1]           \\
     &  0.12   & 0.01747645570 & 0.01748858\footnotemark[2],0.01747645565\footnotemark[3] &
     & 0.0244  & 0.00004715572 & 0.00004715571\footnotemark[3]     \\
     &  0.148  & 0.00009780662 & 0.00009780\footnotemark[3] &
$6d$ &  0.005  & 0.00899421353 & 0.008994\footnotemark[1]                 \\
$3p$ & 0.01    & 0.04561104138 & 0.045611\footnotemark[1]   &
     & 0.0171  & 0.00006413769 &                       \\
     & 0.06    & 0.00447257513 & 0.00447257511\footnotemark[3],0.004472\footnotemark[4]  &
$4f$ & 0.008   & 0.02330635503 &                       \\
     & 0.068   & 0.00030452301 &                               &
     & 0.0352  & 0.00002734812 & 0.0000273481\footnotemark[3] \\
$4p$ & 0.03    & 0.00503284729 & 0.005033\footnotemark[4]      & 
$5f$ & 0.008   & 0.01216554372 &                   \\
     & 0.039   & 0.00010681462 &                               &
     & 0.0234  & 0.00004667366 &                     \\
$5p$ & 0.01    & 0.01040587771 & 0.010401\footnotemark[1]     & 
$6f$ & 0.008   & 0.00624761715 &               \\
     & 0.025   & 0.00012797171 &                               &
     & 0.0167  & 0.00000418053 & 0.00000418053\footnotemark[3]  \\
$6p$ & 0.01    & 0.00467631494 & 0.004651\footnotemark[1]     &
$5g$ & 0.005   & 0.01503173481 & 0.015032\footnotemark[1]           \\
     & 0.017   & 0.00027324316 &                              &    
     & 0.0223  & 0.00004909577 &                      \\
$3d$ & 0.005   & 0.05056063169 & 0.050561\footnotemark[1]          &  
$6g$ & 0.005   & 0.00896866219 & 0.008967\footnotemark[1] \\
     & 0.0635  & 0.00005036825 & 0.0000503682\footnotemark[3]     &  
     & 0.016   & 0.00005984697 &                           \\
$4d$ & 0.005   & 0.02626968434 & 0.026270\footnotemark[1]         &  
$6h$ & 0.005   & 0.00895010682 & 0.008950\footnotemark[1]  \\
     & 0.0374  & 0.00000260255 & 0.00000260256\footnotemark[3]    &  
     & 0.0154  & 0.00003895032 &                        \\
\end{tabular}
\end{ruledtabular}
\begin{tabbing}
$^{\mathrm{a}}$Ref.~\cite{lam72}. \hspace{100pt}  \=
$^{\mathrm{b}}$Ref.~\cite{ikhdair93}. \hspace{100pt}  \=
$^{\mathrm{c}}$Ref.~\cite{nasser11}. \hspace{100pt}  \=
$^{\mathrm{d}}$Ref.~\cite{lai82}. \hspace{100pt}  \=
\end{tabbing}
\end{table}
\endgroup

Next in Table II, we report all the $l \ne 0$ states belonging to $n=2-6$, at selected values of 
$\delta$. Here also the $\delta$ values are chosen so as to reflect both weak and strong couplings 
of the interaction, with respective $\delta_c$ values mentioned at the top of table. While the 
literature results are clearly quite scanty in comparison to $l=0$ case in Table I, wherever available, 
these are quoted
appropriately. Once again, the existing best result is apparently the one from J-matrix calculation 
\cite{nasser11}. As seen before, in all these cases again, our energy values are virtually identical to 
this method. And whenever
these are not available, the present work produces quite superior eigenvalues compared to the other 
existing values. Note that, for all these eigenstates, we have chosen at least one $\delta$
value, which is very close to the $\delta_c$ value, as difficulties are encountered in these areas, 
with some of the methods. Near the critical values of $\delta$, in general, one needs to extend 
$r_{max}$ to some larger values. For example, 
for a converged result for $4d$ state at $\delta=0.0374$, an $r_{max}$ of 1100 a.u., was employed. This,  
again, is similar to a situation we came across for higher states in the previous paragraph, and also
for some other central potentials \cite{roy05a}.

\begin{figure}
\begin{minipage}[c]{0.40\textwidth}
\centering
\includegraphics[scale=0.45]{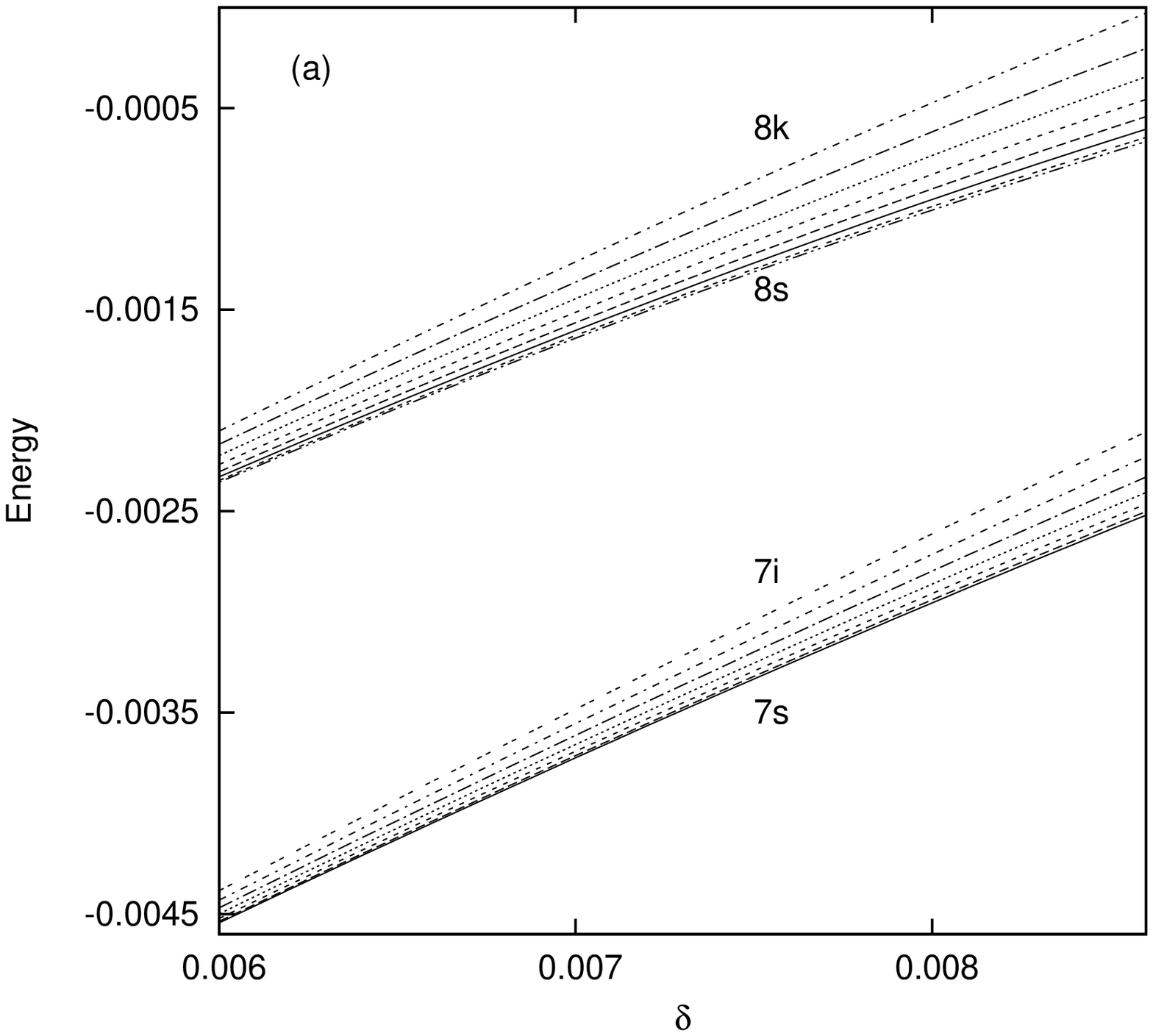}
\end{minipage}%
\hspace{0.5in}
\begin{minipage}[c]{0.40\textwidth}
\centering
\includegraphics[scale=0.45]{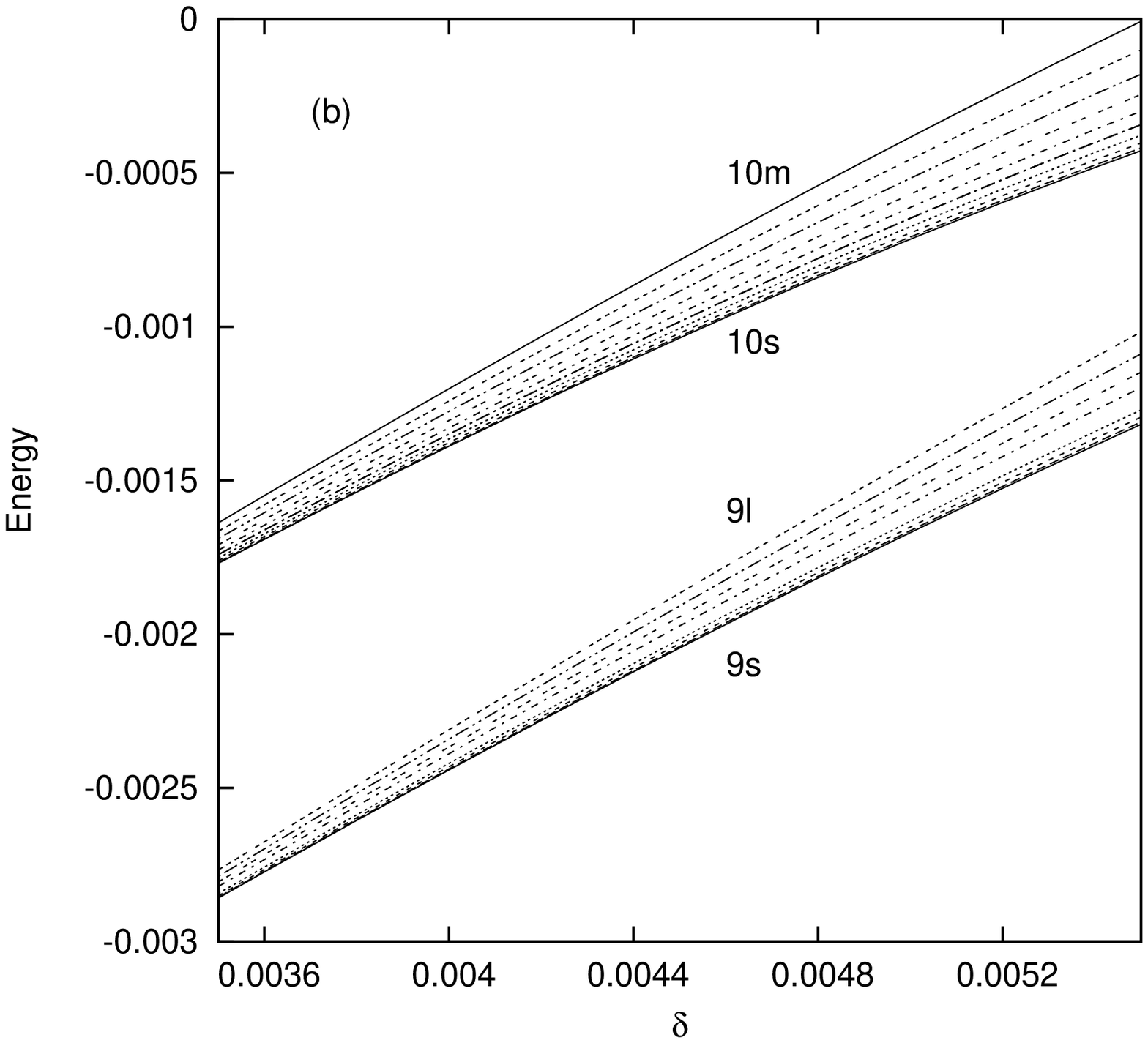}
\end{minipage}%
\caption{Energy eigenvalues (a.u.) of the ECSC potential for (a) $n=7,8$ and (b) $n=9,10$ levels 
respectively as a function of $\delta$ in the vicinity of zero energy.}
\end{figure}

Next in Fig.~1, variation of energy eigenvalues with respect to screening parameters are depicted for 
all the states belonging to $n=7,8$ (left) and $n=9,10$ (right) respectively, in the neighborhood of
zero energy. Energy values increase monotonically with $\delta$; for each $n$, they make a distinct
family and for a particular value of the quantum number $n$, the separation between states with different
values of $l$ tends to increase with an increase in $\delta$. Additionally, in Table III, calculated 
eigenvalues of all states are given at selected values of $\delta$ (0.005 and 0.003 for $n=8$ and
10 respectively). For sake of completeness, the available $\delta_c$ values for $n=8$ are mentioned in 
column 1 in parentheses. Only a two-parameter variational calculation \cite{lam72} has been reported for 
the $n=8$ state. While these are reasonable first estimates in absence of any other result, our GPS 
results are significantly better than these. And for $n=10$, there are no results to quote for direct 
comparison, and it is hoped that these would be helpful for the purpose of future referencing. 
     
\begingroup
\squeezetable
\begin{table}
\caption {\label{tab:table3}Comparison of the negative eigenvalues (a. u.) of 
ECSC potential for $n =8,10$ states at selected values of $\lambda$. Numbers in the
parentheses in column 1 denote $\delta_c$ values \cite{nasser11}.} 
\begin{ruledtabular}
\begin{tabular}{llllllll}
State & $\delta$ & \multicolumn{2}{c}{$-$Energy} & State &  $\delta$ & 
\multicolumn{2}{c}{$-$Energy}   \\ 
\cline{3-4} \cline{7-8}
   &  & This work & Literature \cite{lam72} &  &   &   This work & Literature \\    \hline 
$8s$(0.0100)  & 0.005 & 0.00314139349 & 0.003134 & $10s$ & 0.003 & 0.00217587556 &   \\
$8p$(0.0099)  &       & 0.00313602921 & 0.003128 & $10p$ &       & 0.00217402019 &   \\
$8d$(0.0098)  &       & 0.00312524768 & 0.003118 & $10d$ &       & 0.00217029874 &   \\
$8f$(0.0096)  &       & 0.00310894203 & 0.003102 & $10f$ &       & 0.00216468974 &   \\
$8g$(0.0094)  &       & 0.00308694949 & 0.003080 & $10g$ &       & 0.00215716069 &   \\
$8h$(0.0092)  &       & 0.00305904829 & 0.003053 & $10h$ &       & 0.00214766774 &   \\
$8i$(0.0089)  &       & 0.00302495321 & 0.003020 & $10i$ &       & 0.00213615531 &   \\
$8k$(0.0086)  &       & 0.00298430926 & 0.002981 & $10k$ &       & 0.00212255542 &   \\
              &       &               &          & $10l$ &       & 0.00210678692 &   \\
              &       &               &          & $10m$ &       & 0.00208875461 &   \\
\end{tabular}
\end{ruledtabular}
\end{table}
\endgroup

Now we turn to the GESC potential. Table IV reports energies for some low- and high-lying $s$ states of the 
same for some representative $b$ values, keeping $a$ fixed at 1. For lower states, several $b$ values are 
considered to understand the dependence on potential parameters. Reference results are much scarce in this case 
compared to the ECSC potential. The $1/N$-expansion up to 14 terms \cite{sever87a} were obtained for ground 
and first excited state energies and wave functions. While these are reasonable initial estimates, 
clearly improved energies would be highly desirable. In another treatment, bound-state energies of first 
three $s$ states have been reported within a new perturbation technique \cite{ikhdair07a}. For smaller
screening parameters, there is, in general, a decent agreement between our result and theirs. 
However, the discrepancy starts to grow quite fast as $b$ is increased. Finally, in Table V, eigenvalues of all 
the $l \ne 0$ states having $n \le 8$ are reported for the first time, for two values of $b$ parameter. 
No results could be found for such states in the literature for comparison. This dependence of GESC 
eigenvalues on $b$ is pictorially shown in Fig.~2 for all the states belonging to $n=2,3$ (a), 
$n=4,5$ (b), $n=7,8$ (c) and $n=9,10$ (d) respectively. In all cases, energies gradually increase and then 
tend to assume a constant value. For smaller $n$ in (a), (b), it is seen that, all the states belonging 
to a particular $n$ form a characteristic family of the curve. Moreover, the states corresponding to a 
given $n$ do not mix with the states with a different $n$. However, this scenario changes dramatically 
as we go for higher $n$. Thus, as we move to (c), appreciable complex ordering and inter-state mixing is 
observed for $n=6,7$ ($7i$ mixing with $8s$) and as we finally reach $n=9,10$ in (d), we encounter 
heavy mixing among the $9i, 9k, 9l$ and $10s, 10p, 10d$ states at around $b=0.03-0.06$, making accurate 
determinate of these eigenvalues more and more difficult. Such complex level crossings have also been 
observed earlier for Hulth\'en and Yukawa potentials \cite{roy05a}. 

\begingroup
\squeezetable
\begin{table}
\caption {\label{tab:table4}Comparison of the negative eigenvalues (a. u.) of 
GESC potential for several $s$ states at selected values of $b$.} 
\begin{ruledtabular}
\begin{tabular}{lllllll}
State  & $b$ & \multicolumn{2}{c}{$-$Energy} &  $b$    &  \multicolumn{2}{c}{$-$Energy}   \\ 
\cline{3-4}  \cline{6-7}
   &  & This work & Literature &  &     This work & Literature \\    \hline 
 $1s$ & 0.001   & 1.99900000049  &                                                    & 
        0.005 & 1.99500006211 &                 \\  
      & 0.02    & 1.98000390178  & 1.9800039\footnotemark[1],1.98000\footnotemark[2]  & 
        0.05  & 1.95005876574 & 1.9500586\footnotemark[1]  \\ 
      & 0.08    & 1.92023217638  & 1.9202305\footnotemark[1],1.92023\footnotemark[2]  & 
        0.2   & 1.80316184099 & 1.8030143\footnotemark[1]  \\
      & 0.4     & 1.62057014563  & 1.6169173\footnotemark[1]                          & 
        0.7   & 1.38422179244 
                                 & 1.3477860\footnotemark[1],1.384\footnotemark[2]      \\
      & 1       & 1.19419978389  & 1.0989583\footnotemark[1],1.194\footnotemark[2]    & 
        2     & 0.82070036307 &                            \\
      & 3       & 0.66846103237  &                                                    & 
        5     & 0.56680152293 &                            \\
      & 10      & 0.51787565892  &                                                    & 
        20    & 0.50467744871 &                            \\
 $2s$ & 0.001   & 0.49900000697  &                                                    & 
        0.005 & 0.49500085803 &                            \\  
      & 0.02    & 0.48005182598  & 0.4800516\footnotemark[1],0.48000\footnotemark[2]  & 
        0.1   & 0.40487183925 
                                 & 0.4043555\footnotemark[1],0.4048\footnotemark[2]       \\
      & 0.3     & 0.27382625160  & 0.2431595\footnotemark[1],0.274\footnotemark[2]    & 
        0.5   & 0.21294420503 &                            \\
      & 1       & 0.17216986942  &                                                    & 
        5     & 0.13273550025 &                            \\
      & 10      & 0.12719012732  &                                                    & 
        15    & 0.12601285391 &                            \\ 
 $3s$ & 0.001   & 0.22122225644  &                                                    & 
        0.02  & 0.20245702303 & 0.2024526\footnotemark[1]  \\
      & 0.06    & 0.16695326252  & 0.1662097\footnotemark[1]                          & 
        0.2   & 0.09999056593 &                            \\
      & 0.5     & 0.07573455915  &                                                    & 
        1     & 0.06809011302 &                            \\
 $4s$ & 0.001   & 0.12400010645  &                                                    & 
        0.05  & 0.08216008138 &                            \\ 
      & 0.1     & 0.05956143168  &                                                    & 
        0.5   & 0.03910015933 &                            \\
 $5s$ & 0.001   & 0.07900025672  &                                                    & 
        0.2   & 0.02693147558 &                            \\
 $6s$ & 0.001   & 0.05455608155  &                                                    & 
        0.1   & 0.01998364734 &                            \\
 $9s$ & 0.001   & 0.02369391160  &                                                    & 
        0.1   & 0.00776025524 &                            \\
\end{tabular}
\end{ruledtabular}
\begin{tabbing}
$^{\mathrm{a}}$Ref.~\cite{ikhdair07a}. \hspace{100pt}  \=
$^{\mathrm{b}}$Ref.~\cite{sever87a}. \hspace{100pt} 
\end{tabbing}
\end{table}
\endgroup

\section{conclusion}
Accurate bound states of ECSC and GESC potential have been presented by means of a GPS method. For both
these cases, zero and non-zero angular momentum states are calculated easily with high accuracy. The 
methodology is simple, efficient and, as shown, produces eigenvalues and eigenfunctions of comparable
accuracy to those of the best available methods found in the literature. All the 55 states lying 
with $n \le 10$ for the former, and 36 states with $n \le 8$ for the latter, are calculated up to eleven
significant figures covering wide ranges of interaction. For the former, our results
are superior to all the existing methods except that of the J-matrix formalism, while for the latter potential,
our results surpass the accuracy of all existing methods. A detailed analysis of the variation of energies 
with respect to potential parameters show quite different trends for these two potentials. For higher $n$, 
complex level 
crossing and inter-state mixing has been observed for the GESC potential. Special attention was paid for
the high-lying states and regions of \emph{strong} screening parameters. Many states are presented here 
for the first time. This offers a simple reliable method for the accurate calculation of these 
and other potentials in quantum mechanics. 

\begingroup
\squeezetable
\begin{table}
\caption {\label{tab:table5}Energies (a. u.) of the GESC potential for several $l \ne 0$ states at 
selected values of $b$.} 
\begin{ruledtabular}
\begin{tabular}{lllllll}
State  & $b$ & $-$Energy &  $b$    &  $-$Energy &  $b$   &  $-$Energy   \\ 
\hline 
 $2p$ & 0.001   & 0.49900000498  &    0.2   &   0.32215735500    &  1      &  0.13471500886     \\  
 $3p$ & 0.001   & 0.22122225198  &    0.05  &   0.17482725023    &  0.5    &  0.06796045435     \\  
 $3d$ &         & 0.22122224308  &          &   0.17414406972    &         &  0.05712005483     \\  
 $4p$ & 0.001   & 0.12400009861  &    0.05  &   0.08176332315    &  0.3    &  0.03976138004     \\  
 $4d$ &         & 0.12400008291  &          &   0.08094164047    &         &  0.03494876135     \\  
 $4f$ &         & 0.12400005934  &          &   0.07963728134    &         &  0.03166956721     \\  
 $5p$ & 0.001   & 0.07900024460  &    0.05  &   0.04265374603    &  0.2    &  0.02588418263     \\  
 $5d$ &         & 0.07900022035  &          &   0.04187154574    &         &  0.02383314014     \\  
 $5f$ &         & 0.07900018393  &          &   0.04063035100    &         &  0.02127952212     \\  
 $5g$ &         & 0.07900013528  &          &   0.03883963267    &         &  0.02015250137     \\  
 $6p$ & 0.001   & 0.05455606434  &    0.05  &   0.02479616384    &  0.2    &  0.01712467704     \\  
 $6d$ &         & 0.05455602989  &          &   0.02418563145    &         &  0.01601163848     \\  
 $6f$ &         & 0.05455597816  &          &   0.02322582411    &         &  0.01465426827     \\  
 $6g$ &         & 0.05455590906  &          &   0.02185943895    &         &  0.01401444073     \\  
 $6h$ &         & 0.05455582250  &          &   0.01999675997    &         &  0.01389838294     \\  
 $7p$ &  0.001  & 0.03981726562  &    0.02  &   0.02456106851    &  0.1    &  0.01354669295     \\  
 $7d$ &         & 0.03981721949  &          &   0.02443236181    &         &  0.01300727690     \\  
 $7f$ &         & 0.03981715020  &          &   0.02423479689    &         &  0.01221827987     \\  
 $7g$ &         & 0.03981705766  &          &   0.02396283009    &         &  0.01127101023     \\  
 $7h$ &         & 0.03981694172  &          &   0.02360879174    &         &  0.01052800497     \\  
 $7i$ &         & 0.03981680223  &          &   0.02316253952    &         &  0.01025534957     \\  
 $8p$ &  0.001  & 0.03025158916  &   0.01   &   0.02226047376    &  0.1    &  0.00996391701     \\  
 $8d$ &         & 0.03025153001  &          &   0.02223011222    &         &  0.00961764441     \\  
 $8f$ &         & 0.03025144118  &          &   0.02218403767    &         &  0.00911555429     \\  
 $8g$ &         & 0.03025132253  &          &   0.02212160821    &         &  0.00852094708     \\  
 $8h$ &         & 0.03025117390  &          &   0.02204196197    &         &  0.00805230527     \\  
 $8i$ &         & 0.03025099505  &          &   0.02194401054    &         &  0.00786434715     \\  
 $8k$ &         & 0.03025078574  &          &   0.02182642900    &         &  0.00781852403     \\  
\end{tabular}
\end{ruledtabular}
\end{table}
\endgroup

\begin{figure}
\centering
\begin{minipage}[t]{0.4\textwidth}\centering
\includegraphics[scale=0.40]{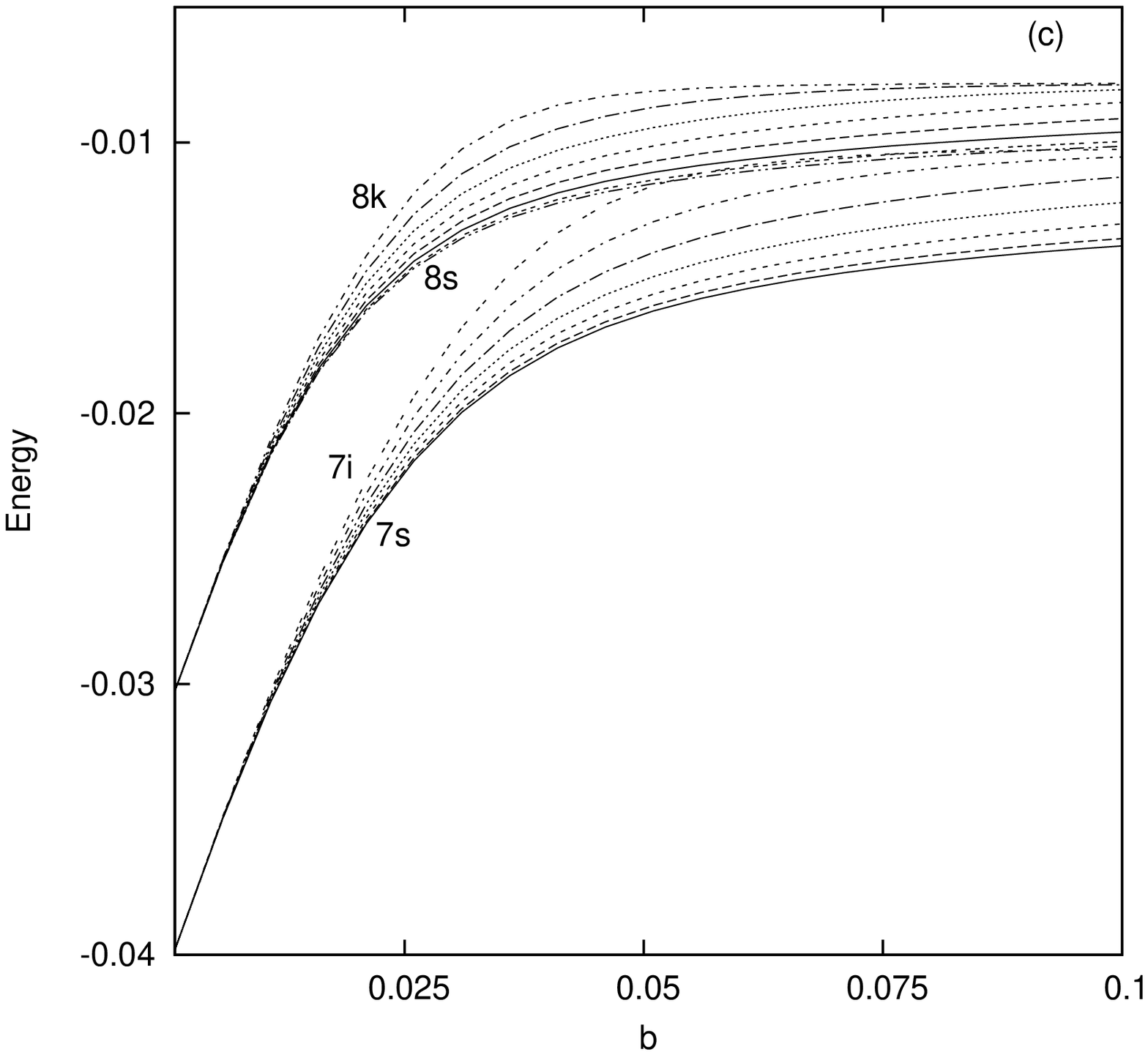}
\end{minipage}
\hspace{0.15in}
\begin{minipage}[t]{0.4\textwidth}\centering
\includegraphics[scale=0.40]{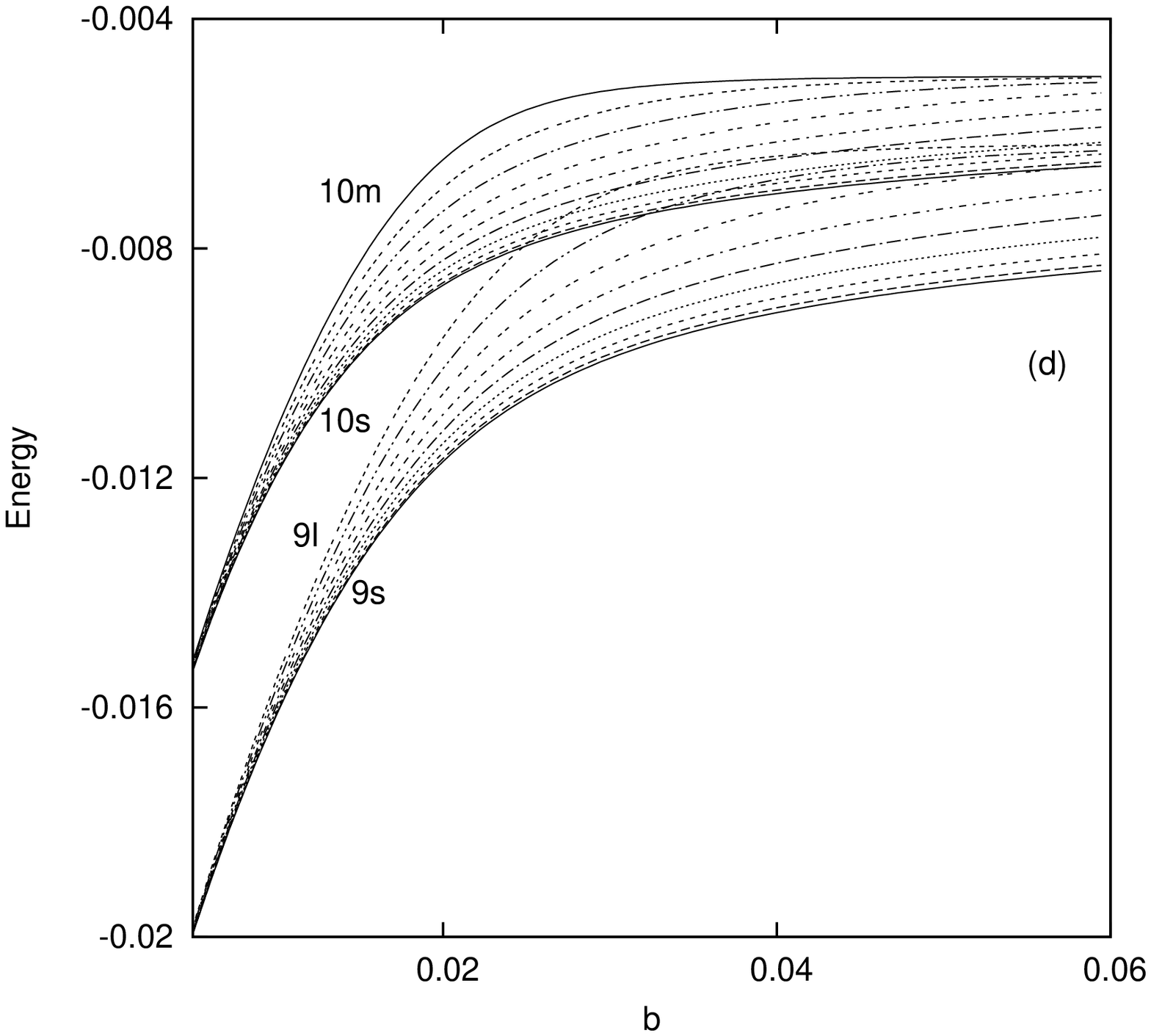}
\end{minipage}
\\[20pt]
\begin{minipage}[b]{0.4\textwidth}\centering
\includegraphics[scale=0.40]{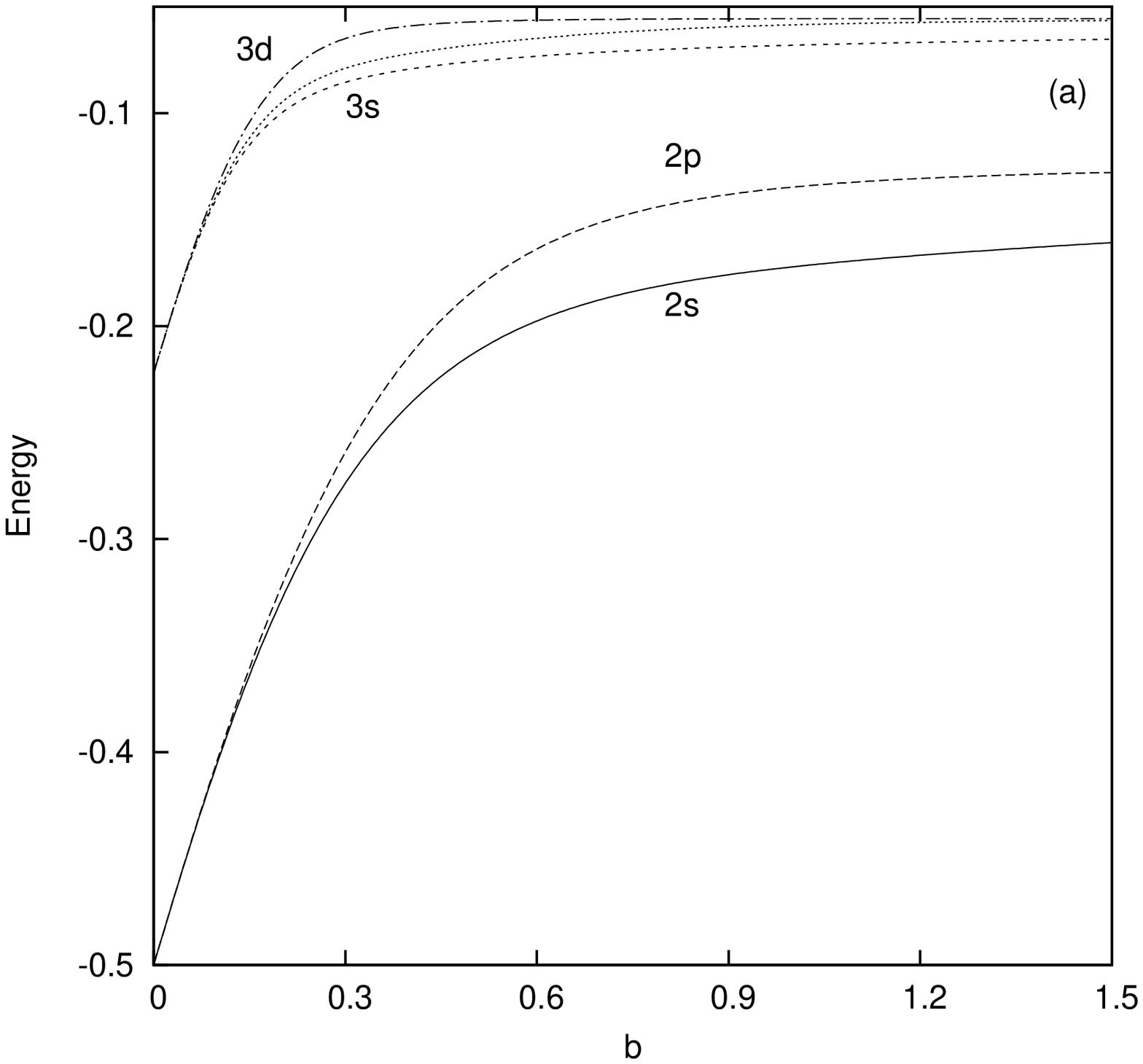}
\end{minipage}
\hspace{0.15in}
\begin{minipage}[b]{0.4\textwidth}\centering
\includegraphics[scale=0.40]{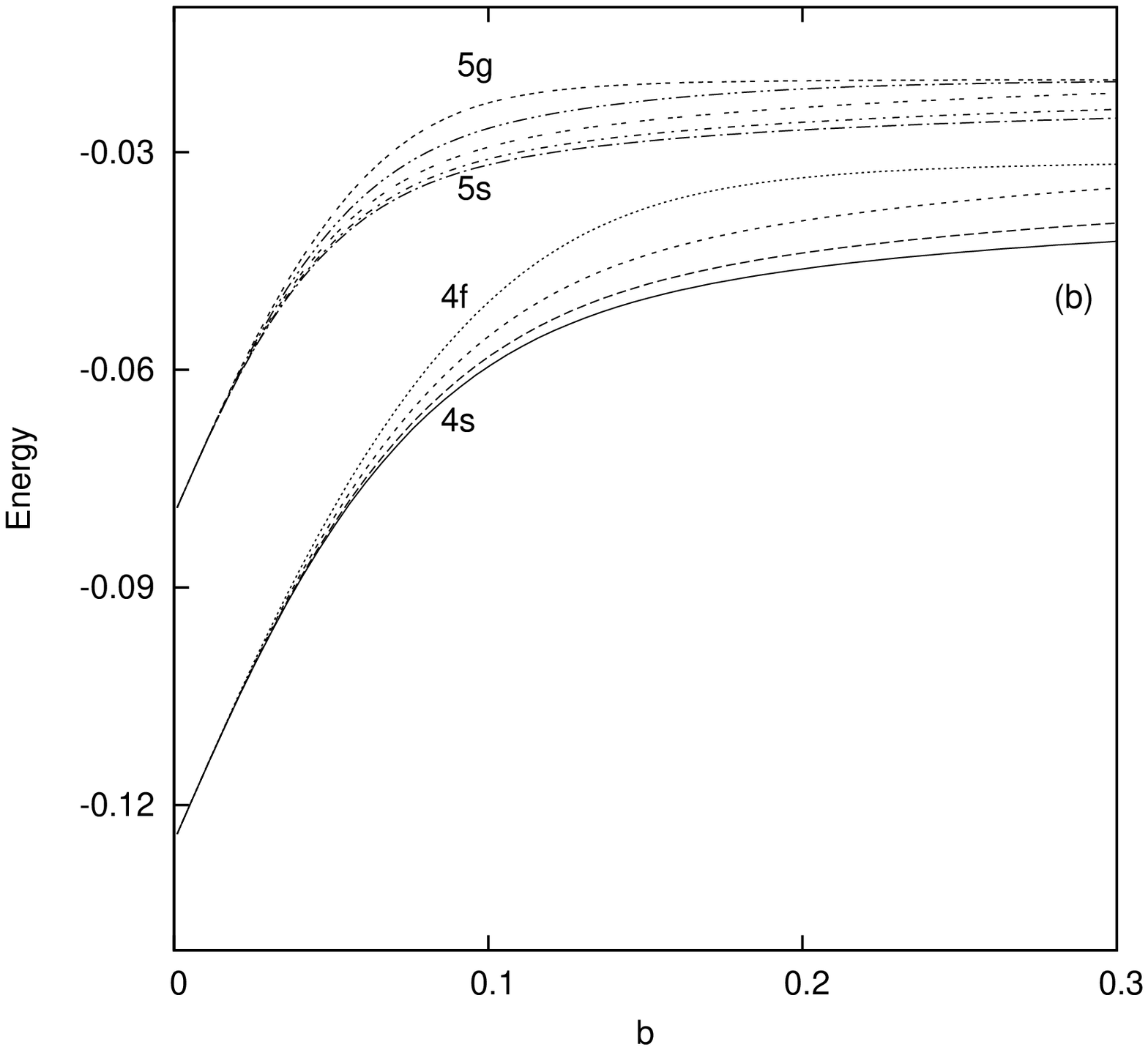}
\end{minipage}
\caption[optional]{Energy eigenvalues (a.u.) of the GESC potential for (a) $n=2,3$ (b) $n=4,5$ 
(c) $n=7,8$ (d) $n=9,10$ levels, respectively, as a function of $b$ in the vicinity of zero energy.}
\end{figure}

\begin{acknowledgments}
The two anonymous referees are thanked for their constructive comments. 
\end{acknowledgments}


\begin{thebibliography}{99}
\bibitem{anderson52} P.~Anderson, Phys.~Rev.~ \textbf{86}, 694 (1952).
\bibitem{kubo52} R.~Kubo, Phys.~Rev.~ \textbf{87}, 568 (1952). 
\bibitem{bonch59} V.~L.~Bonch-Bruevich and V.~B.~Glasko, Sov.~Phys.~Dokl. \textbf{4}, 147 (1959). 
\bibitem{propokev67} E.~P.~Propokev, Sov.~Phys.~Solid State \textbf{9}, 993 (1967). 
\bibitem{ferrell74} R.~A.~Ferrell and D.~J.~Scalapino, Phys.~Rev.~A \textbf{9}, 846 (1974).
\bibitem{brezin79} E.~Brezin, J.~Phys.~A \textbf{12}, 759 (1979). 
\bibitem{weisbuch93} C.~Weisbuch and B.~Vinter, \emph{Quantum Semiconductor Heterostructures}, 
Academic Press, New York (1993).
\bibitem{harrison00} P.~Harrison, \emph{Quantum Wells, Wires and Dots}, John Wiley and Sons, (2000). 
\bibitem{shukla08} P.~K.~Shukla and B.~Eliasson, Phys.~Lett.~A \textbf{372}, 2897 (2008). 
\bibitem{lin10} C.~Y.~Lin and Y.~K.~Ho, Eur.~Phys.~J.~D \textbf{57}, 21 (2010). 
\bibitem{ghoshal09} A.~Ghoshal and Y.~K.~Ho, J.~Phys.~B \textbf{42}, 075002 (2009).
\bibitem{ghoshal09a} A.~Ghoshal and Y.~K.~Ho, Phys.~Rev.~A \textbf{79}, 062514 (2009).
\bibitem{ghoshal11} A.~Ghoshal and Y.~K.~Ho, Int.~J.~Quant.~Chem.~ \textbf{111}, 4288 (2011).
\bibitem{ghoshal11a} A.~Ghoshal and Y.~K.~Ho, mod.~Phys.~Lett.~B \textbf{25}, 1619 (2011).
\bibitem{lam72} C.~S.~Lam and Y.~P.~Varshni, Phys.~Rev.~A \textbf{6}, 1391 (1972). 
\bibitem{dutt79} R.~Dutt, Phys.~Lett.~ \textbf{73A}, 310 (1979). 
\bibitem{ray80} P.~P.~Ray and A.~Ray, Phys.~Lett.~ \textbf{78A}, 443 (1980). 
\bibitem{lai82} C.~S.~Lai, Phys.~Rev.~A \textbf{26}, 2245 (1982).             
\bibitem{meyer85} H.~de Meyer, V.~Fack and G.~Vanden Berghe, J.~Phys.~A \textbf{18}, L849 (1985). 
\bibitem{sever90} R.~Sever and C.~Tezcan, Phys.~Rev.~A \textbf{41}, 5205 (1990). 
\bibitem{singh83} D.~Singh and Y.~P.~Varshni, Phys.~Rev.~A \textbf{28}, 2606 (1983).
\bibitem{sever87} R.~Sever and C.~Tezcan, Phys.~Rev.~A \textbf{35}, 2725 (1987). 
\bibitem{ikhdair93} S.~M.~Ikhdair and R.~Sever, Z.~Phys.~D \textbf{28}, 1 (1993). 
\bibitem{bayrak07} O.~Bayrak and I.~Boztosun, Int.~J.~Quant.~Chem.~ \textbf{107}, 1040 (2007).  
\bibitem{ikhdair07} S.~M.~Ikhdair and R.~Sever, J.~Math.~Chem.~ \textbf{41}, 329 (2007).
\bibitem{nasser11} I.~Nasser, M.~S.~Abdelmonem and Afaf Abdel-Hady, Phys.~Scr.~ \textbf{84}, 
045001 (2011). 
\bibitem{paul11} S.~Paul and Y.~K.~Ho, Computer Phys.~Comm.~ \textbf{182}, 130 (2011). 
\bibitem{bahlouli12} H.~Bahlouli, M.~S.~Abdelmonem and S.~M.~Al-Morzoug, Chem.~Phys.~ 
\textbf{393}, 153 (2012). 
\bibitem{dutt80} R.~Dutt, Phys.~Lett.~ \textbf{77A}, 229 (1980). 
\bibitem{sever87a} R.~Sever and C.~Tezcan, Phys.~Rev.~A \textbf{36}, 1045 (1987). 
\bibitem{ikhdair07a} S.~M.~Ikhdair and R.~Sever, J.~Math.~Chem.~ \textbf{41}, 343 (2007).
\bibitem{roy04} A.~K.~Roy, Phys.~Lett.~A \textbf{321}, 231 (2004).
\bibitem{roy04b} A.~K.~Roy, J.~Phys.~B \textbf{37}, 4369 (2004); {\it ibid.} \textbf{38}, 1591 (2005).
\bibitem{roy05} A.~K.~Roy, Int.~J.~Quant.~Chem.~\textbf{104}, 861 (2005).
\bibitem{roy05a} A.~K.~Roy, Pramana--J.~Phys.~ \textbf{65}, 01 (2005). 
\bibitem{roy07} A.~K.~Roy and A.~F.~Jalbout, Chem.~Phys.~Lett.~ \textbf{445}, 355 (2007). 
\bibitem{roy08} A.~K.~Roy, A.~F.~Jalbout and E.~I.~Proynov, Int.~J.~Quant.~Chem.~ \textbf{108}, 
827 (2008). 
\bibitem{roy08a} A.~K.~Roy, A.~F.~Jalbout and E.~I.~Proynov, J.~Math.~Chem.~ \textbf{44}, 260 (2008). 
\bibitem{roy11} A.~K.~Roy, in \emph{Mathematical Chemistry}, W.~I~Hong (Ed.), Nova Science Publishers, 
Hauppauge, NY, USA, pp.~555-599 (2011). 
\bibitem{yao93} G.~Yao and S.~I.~Chu, Chem.~Phys.~Lett.~ \textbf{204}, 381 (1993). 
\bibitem{wang94} J.~Wang, S.~I.~Chu and C.~Laughlin, Phys.~Rev.~A \textbf{50}, 3208 (1994). 
\bibitem{telnov99} D.~A.~Telnov and S.~I.~Chu, Phys.~Rev.~A \textbf{59}, 2864 (1999). 
\end{thebibliography}
\end{document}